\documentclass[paper]{elsarticle}

\usepackage{amsfonts,amsbsy,amssymb,amsmath}
\usepackage{hyperref}
\usepackage{graphicx}
\usepackage[table,xcdraw]{xcolor}
\usepackage{float}

\graphicspath{{figures/}}

\journal{Chemical Physics Letters}

\biboptions{compress}

\begin{document}

\begin{frontmatter}

\title{The Influence of a Pitchfork Bifurcation of the Critical Points of a Symmetric Caldera Potential Energy Surface on Dynamical Matching}


 \author[label1]{Y. Geng}
 \author[label1]{M. Katsanikas}
 \author[label1]{M. Agaoglou}
 \author[label1]{S. Wiggins\corref{mycorrespondingauthor}}
 \ead{S.Wiggins@bristol.ac.uk}

  \address[label1]{School of Mathematics, University of Bristol, \\ Fry Building, Woodland Road, Bristol, BS8 1UG, United Kingdom.\\[.2cm]}

\begin{abstract}
Many organic chemical reactions are governed by potential energy surfaces that have a region with the topographical features of a caldera. If the caldera has a symmetry then trajectories transiting  the caldera region are observed to exhibit a phenomenon that is referred to as dynamical matching. Dynamical matching is a constraint that restricts the way in which a trajectory can exit the caldera based solely on how it enters the caldera. In this paper we show that bifurcations of the critical points of the caldera  potential energy surface can destroy dynamical matching even when the symmetry of the caldera is not affected by the bifurcation.  
\end{abstract}


\end{frontmatter}


\section{Introduction}
In this paper we study a symmetric caldera potential energy surface (PES) and the effect of bifurcation of critical points. The geometry of this PES resembles that of the collapsed region  of an erupted volcano.  This feature was the main reason Doering \cite{doering2002} and co-workers  used the word ‘’caldera’’ in reference to this type of PES. Features of the caldera PES occur in  many organic chemical reactions,  such as  the   vinylcyclopropane-cyclopentene rearrangement \cite{baldwin2003,gold1988}, the stereomutation of cyclopropane \cite{doubleday1997}, the degenerate rearrangement of bicyclo[3.1.0]hex-2-ene \cite{doubleday1999,doubleday2006} or that of 5-methylenebicyclo[2.1.0]pentane \cite{reyes2002}. 

The topography of the caldera PES  is characterized by a shallow minimum and four index-1 saddles that surround this region.  Two of the index-1 saddles have higher energy (upper index-1 saddles) than the other two (lower index-1 saddles). The four index-1 saddles control the entrance into and  exit from the central area of the caldera. Chemical  reaction occurs when trajectories from the region of one of the upper index-1 saddles (reactants) cross the caldera and approach the region of one of the two lower index-1 saddles (products). 

Detailed studies of trajectories in a two dimensional caldera PES have been carried out in \cite{collins2014,katsanikas2018,katsanikas2019,katsanikas2020a,katsanikas2020b}).  Broadly speaking, there are two distinct situations: the case when the PES is symmetric and the case when the PES is asymmetric (to be more precisely defined below). The symmetric case was studied in \cite{collins2014,katsanikas2018} where it was found that trajectories entering the caldera from the region of an upper index one saddle exited the caldera from the {\em opposite} lower index one saddle. Hence $100\%$ of the products exit from the region of {\em one}  lower index-1 saddle and not $50\%$, as  would be predicted by statistical theories. This phenomenon is referred to as {\em dynamical matching} and was first reported in  (\cite{carpenter1985,carpenter1995}). 

A phase space analysis of dynamical matching was carried out in \cite{katsanikas2018}) where it was shown that the controlling mechanism is the existence of heteroclinic orbits between the unstable manifolds of the unstable periodic orbits associated with the upper saddles and the stable manifolds of invariant sets in the central region. In particular, the non-existence of such heteroclinic orbits implies the existence of dynamical matching and the existence of such heteroclinic orbits promotes trapping in the central region and inhibits dynamical matching. In all cases of the symmetric caldera PES studied thus far the non-existence of such heteroclinic trajectories was found and, hence, dynamical matching was always observed. In \cite{collins2014, katsanikas2019} it was shown that if the symmetry of the caldera PES was broken by ‘’stretching’’ the PES then dynamical matching could be broken. Indeed, \cite{katsanikas2019,katsanikas2020a,katsanikas2020b}) showed that a rich variety of dynamical behavior could be created as a result.

In this paper we show that there is a different mechanism for breaking dynamical matching in a symmetric caldera PES that does {\em not } require symmetry breaking.  Rather, it involves a pitchfork bifurcation of the critical points of the PES.

 The structure of the paper is as follows: In  section \ref{model} we describe the structure of the caldera PES. In section \ref{sec:bifPES} we describe the pitchfork bifurcation of the critical points of the PES and compare the topography of the PES before and after the bifurcation. 
In  section \ref{results} we show how the bifurcation influences trajectories crossing the caldera and demonstrate its effect on dynamical matching. In section \ref{conclusions} we summarize our conclusions.

\section{Model}
\label{model}

In this section we describe the Caldera potential energy surface (PES) and its geometrical features that will be the focus of our dynamical studies. The caldera PES that we use is taken from \cite{collins2014} and has the form:

\begin{equation}
    \begin{aligned}
    V(x,y) &= c_1r^2+c_2y-c_3r^4\cos(4\theta)\\ &= c_1(x^2+y^2)+c_2y-c_3(x^4+y^4-6x^2y^2),
    \end{aligned}
\end{equation}

\noindent
where $(x,y)$  are cartesian coordinates, $(r,\theta)$  are standard polar  coordinates, and 
$c_1,c_2,c_3$ are parameters. We make the important observation that the PES is 
symmetric with respect to the y-axis.

Hamilton’s equations are straightforward. We let $p_x$ and $p_y$ denote the momentum of the particle in $x-$ and $y-$ direction respectively, the 2 degree-of-freedom (DoF) Hamiltonian is given by:

\begin{equation}
    H(x,y,p_x,p_y)=\frac{p_x^2}{2m}+\frac{p_y^2}{2m}+V(x,y),
\end{equation}

\noindent
where we consider $m$ to be a constant equal to $1$. The Hamiltonian equations of motion are therefore:

\begin{equation}
    \begin{aligned}
    &\dot{x} = \frac{\partial H}{\partial p_x} = \frac{p_x}{m}\\
    &\dot{y} = \frac{\partial H}{\partial p_y} = \frac{p_y}{m}\\
    &\dot{p_x} = - \frac{\partial V}{\partial x}(x,y) = -(2c_1x-4c_3x^3+12c_3xy^2)\\
    &\dot{p_y} = - \frac{\partial V}{\partial y}(x,y) = -(2c_1y-4c_3y^3+12c_3x^2y+c_2)
    \end{aligned}
\end{equation}

\section{Bifurcation of the Critical Points of the Caldera PES}
\label{sec:bifPES}

In  (\cite{collins2014,katsanikas2018,katsanikas2019,katsanikas2020a,katsanikas2020b}) the parameters of the  caldera PES were fixed at $c_1=5,c_2=3$ and $c_3=-0.3$. In this case the PES  has one minimum  and with four index-1 saddles, two for low values of energy and other two for high values of energy, as listed in table \ref{table:1}. The index-1 saddles control the exit from and the entrance into the caldera. 

We now consider how the critical points of the PES change as we vary $c_1$ but leave the other two parameters fixed at $c_2=3$ and $c_3=-0.3$. It is important to note that the PES remains symmetric, in the sense described above, for all range of values of $c_1$ that we consider.

We consider $c_1$ in the interval $0$ to $5$ and find that there is a critical value of $c_1=1.32$   for which the number of critical points of the PES change from 3 to 5. This is a result of a pitchfork bifurcation involving the minimum and the two lower energy index-1 saddles. Hence, for $0   \le c_1 \le 1.32$ the two higher energy index-1 saddles exist and one lower energy index-1 saddle on the $y$-axis. In other words, there is no longer a minimum which is indicative of the existence of a well. This is a regime for the symmetric caldera which has not received attention from the point of view of trajectory analysis.

In table \ref{table:2}    we give the location of the critical points and their energies for $c_1=0.4$.

\begin{table}
\begin{center}
\caption{Stationary points of the caldera potential for $c_1=0.4,c_2=3$ and $c_3=-0.3$  ("RH" and "LH" are the abbreviations for right hand and left hand respectively)}
\label{table:2}
\begin{tabular}{l  l  l  l}
\hline
Critical point & x & y & E \\
\hline
Lower  saddle & 0.000 & -1.194 & -2.402 \\
Upper LH saddle  & -1.204 & 0.840  & 2.321 \\
Upper RH saddle  & 1.204 &  0.840 & 2.321 \\
\hline
\end{tabular}
\end{center}
\end{table}

In table \ref{table:1}    we give the location of the critical points and their energies for $c_1=5$.

\begin{table}
\begin{center}
\caption{Stationary points of the caldera potential for $c_1=5,c_2=3$ and $c_3=-0.3$  ("RH" and "LH" are the abbreviations for right hand and left hand respectively)}
\label{table:1}
\begin{tabular}{l  l  l  l}
\hline
Critical point & x & y & E \\
\hline
Central minimum & 0.000 & -0.297 & -0.448 \\
Upper LH saddle  & -2.149 & 2.0778 & 27.0123 \\
Upper RH saddle  & 2.149 &  2.0778 & 27.0123 \\
Lower LH saddle & -1.923 & -2.003  & 14.767 \\
Lower RH saddle & 1.923 & -2.003 & 14.767 \\
\hline
\end{tabular}
\end{center}
\end{table}

\begin{figure}
 \centering
A)\includegraphics[scale=0.55]{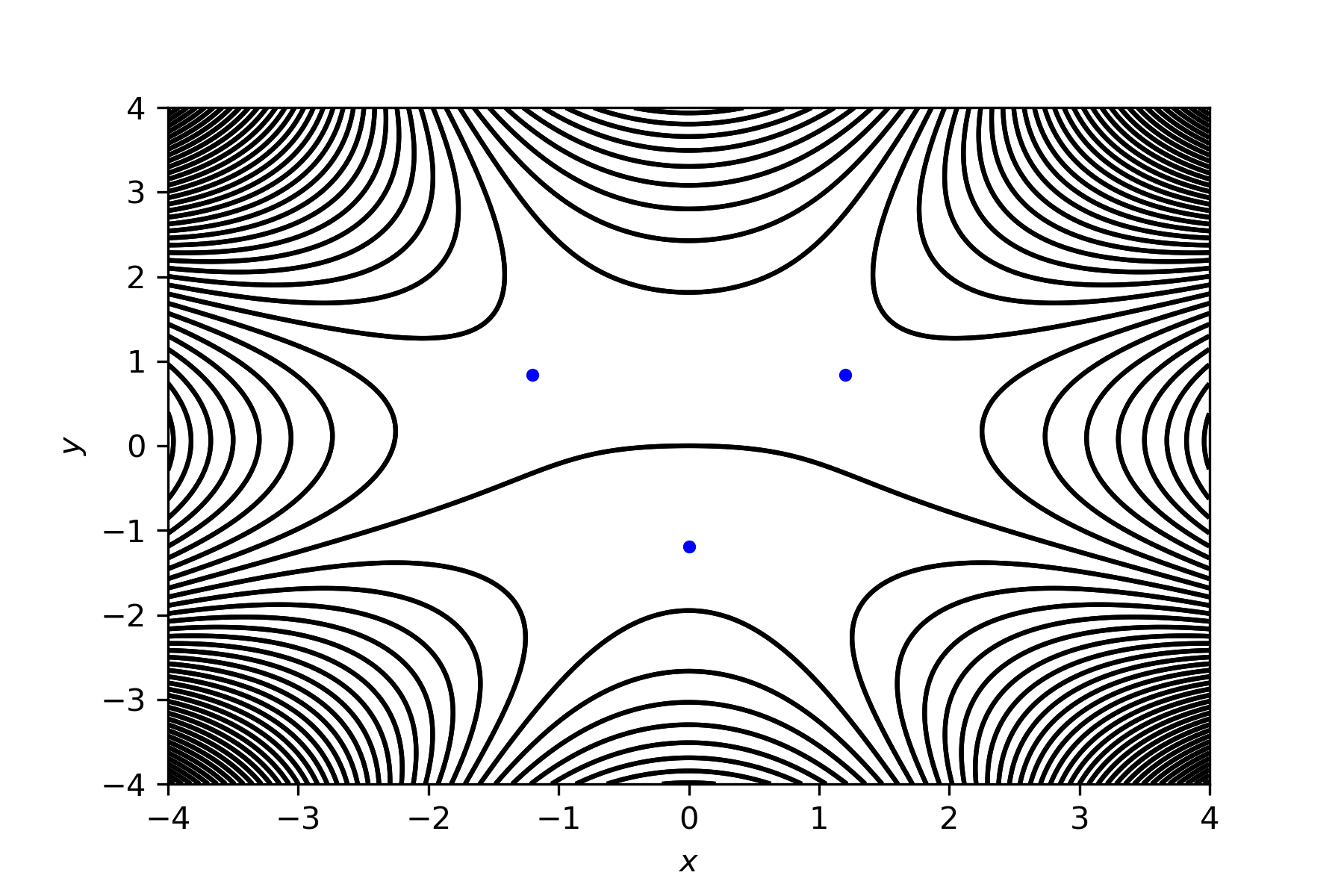}
B)\includegraphics[scale=0.55]{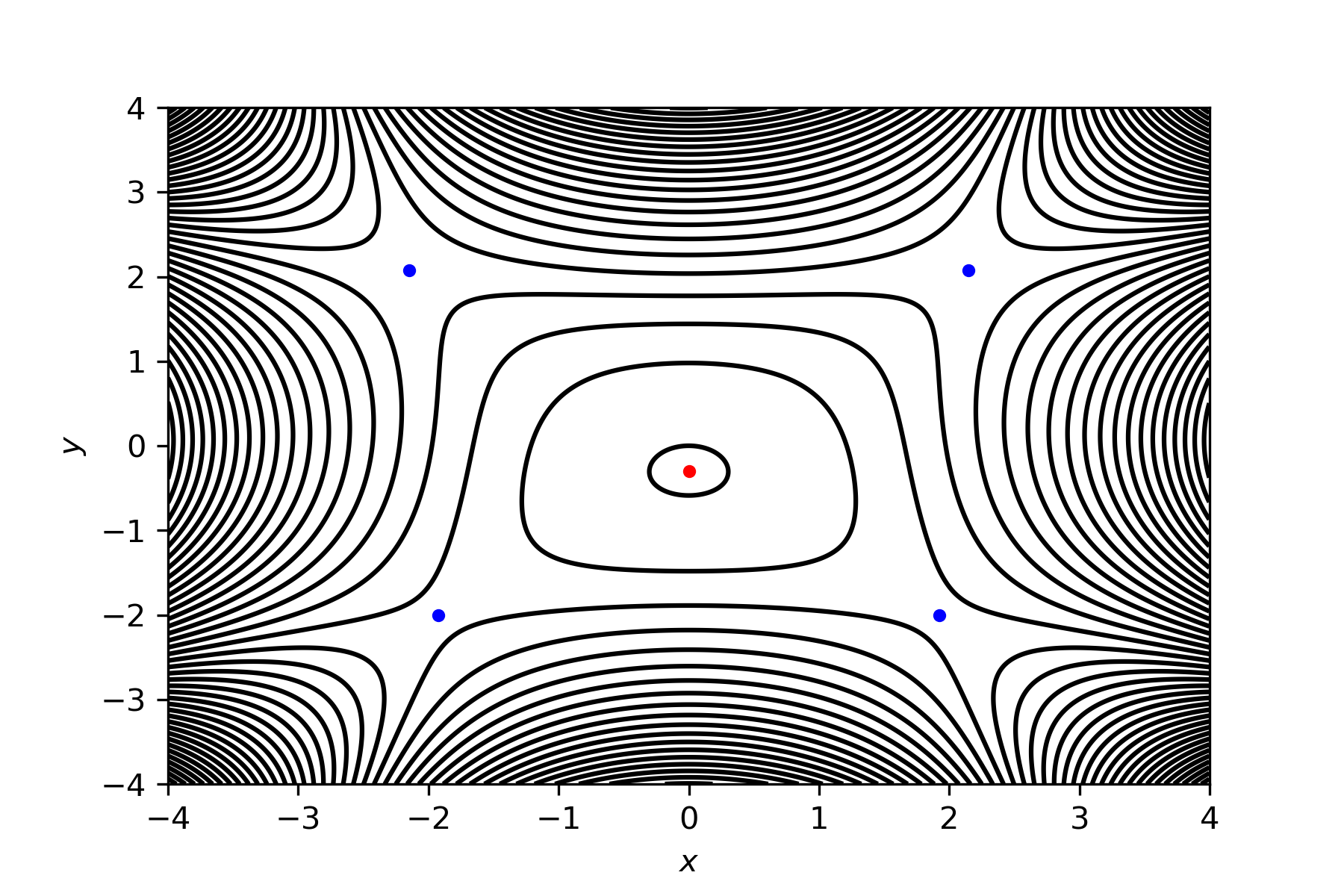}\\
\caption{The contours of the caldera potential: A) for $c_1=0.4,c_2=3$ and $c_3=-0.3$. B)  for $c_1=5,c_2=3$ and $c_3=-0.3$. We indicate the location of the index-1 saddles and of the centers using blue and red points, respectively.}
\label{contours}
\end{figure}

In Fig. \ref{contours} we show contours of the PES for two sets of parameter values showing the two different configurations of critical points, i.e. 3 critical points in panel A and 5 critical points in panel B.

\begin{figure}
 \centering
A)\includegraphics[scale=0.15]{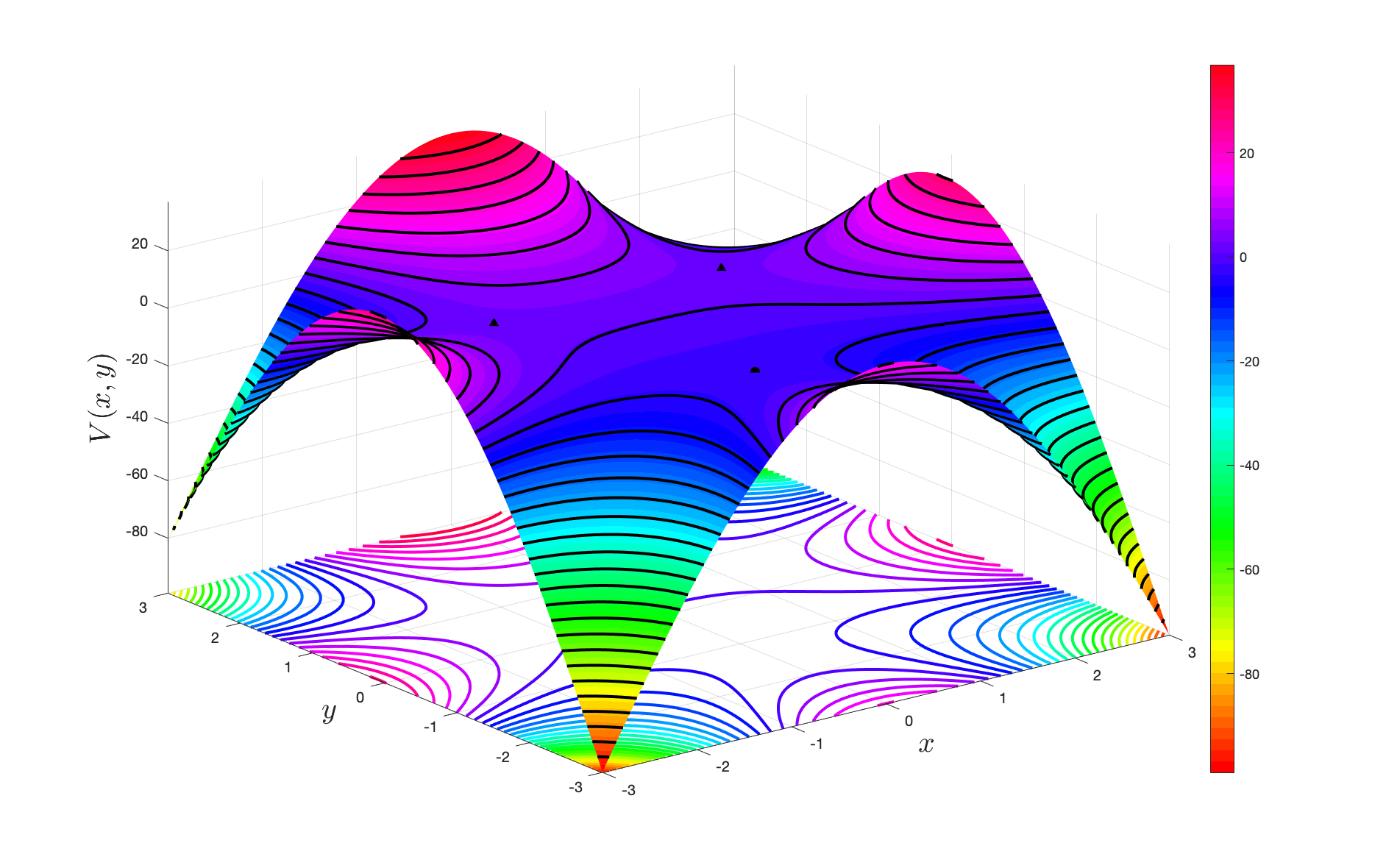}
B)\includegraphics[scale=0.15]{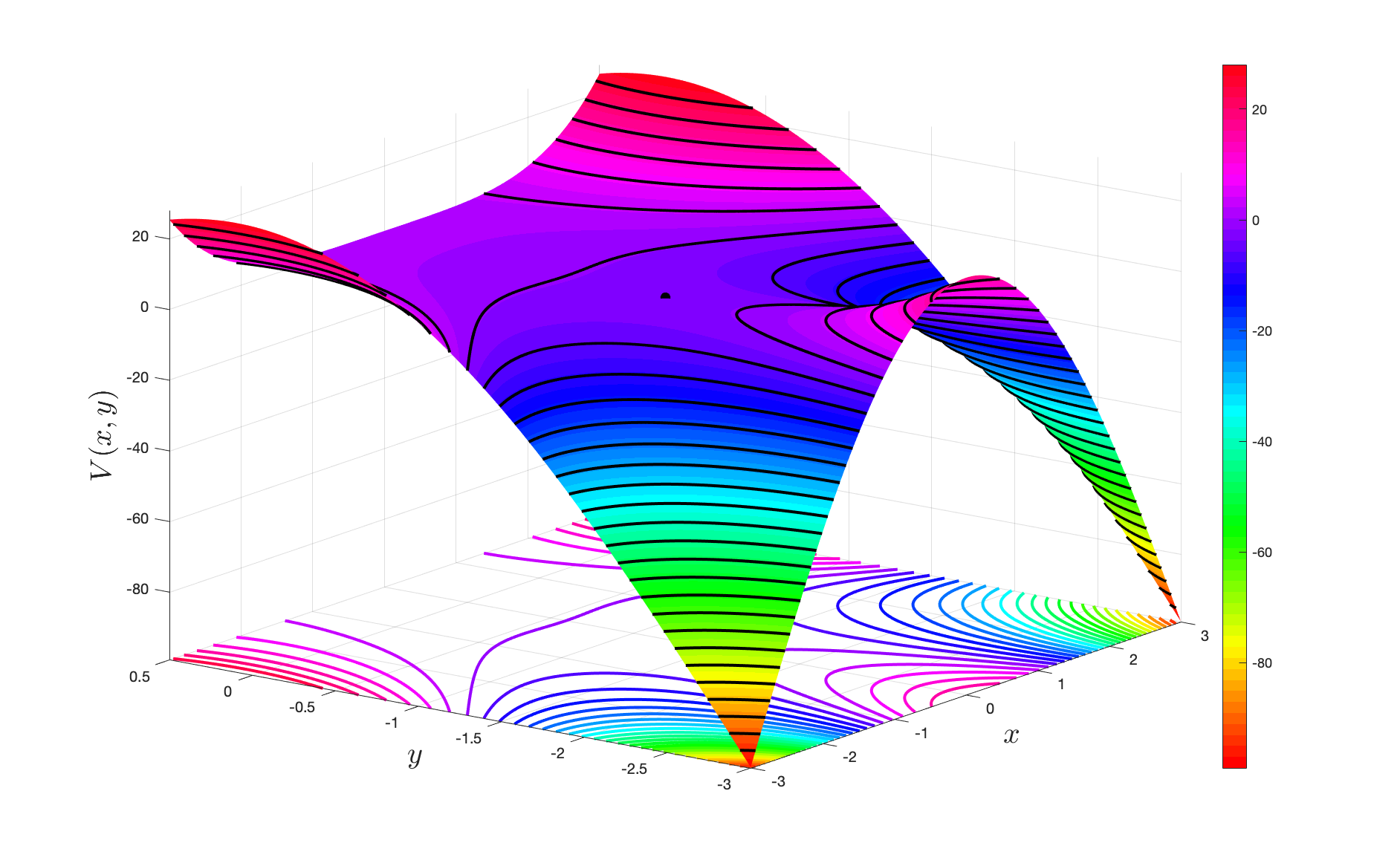}\\
C)\includegraphics[scale=0.15]{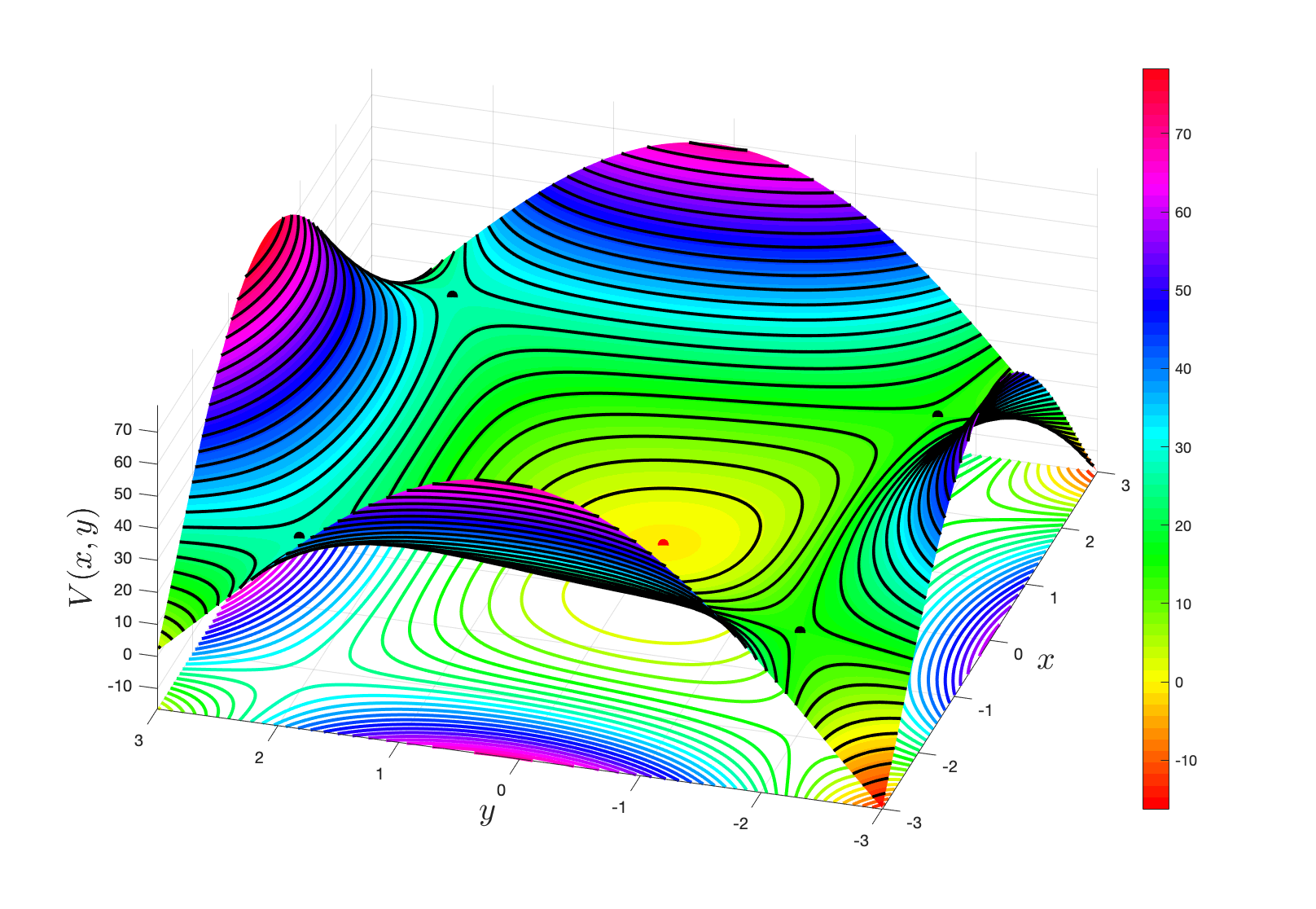}\\
\caption{The 3D PES of the caldera potential A) and B) for  $c_1=0.4,c_2=3$ and $c_3=-0.3$. C) for $c_1=5,c_2=3$ and $c_3=-0.3$. We indicate the location of the index-1 saddles and of the center using black and red points respectively.}
\label{3Dcontours}
\end{figure}

In Fig. \ref{3Dcontours} we show 3D views of the PES for the same set of parameter values (panels A and B show different views for the same set of parameter values).

\section{Results}
\label{results}

In this section we investigate the behavior  of  trajectories that enter the caldera from the region of the upper index-1 saddles for  $0 \le c_1 \le 5$. To achieve this we will analyze the fate of the trajectories that are initialized in the region of the right upper index-1 saddle. The results will be similar if we choose the other upper index-1 saddle due to the symmetry of the potential. 

First, we specify the choice of initial conditions. In particular, we choose a line in configuration space that passes through the upper right index-1 saddle and  is perpendicular to the line that connects the upper right index-1 saddle with the lower index-1 saddle on the y-axis for $0 \le c_1 \le 1.32$   or the lower left index-1 saddle  for $1.32 \le c_1 \le 5$.  The momentum for these initial conditions is chosen to be in the direction of  the line that connects the two index-1 saddles and to satisfy the constant energy condition. This choice guarantees that the initial conditions correspond to trajectories that enter into the caldera from the region of the upper right index-1 saddle. 

We choose 1000 initial conditions in this way (equally spaced along the chosen line in configuration space) and for each initial conditions $c_1$ is chosen  in increments of $0.01$ in the appropriate interval. These initial conditions are integrated for a fixed time interval $t=3$ time units.

We consider that a trajectory has exited through the lower left exit region  or the lower right exit region, if  the $y$ component of the  trajectories are below the line $y=-2.5$ and it has a  negative $x$-coordinate or a positive $x$-coordinates, respectively. We note that the lower left exit region and the lower right exit region are the regions on the left, or on the right, of the  index-1 saddle of the $y$-axis, respectively, before the bifurcation, i.e. for $0 \le c_1 \le 1.32$.   

Similarly, the lower left exit region and the lower right exit region are the region of the lower left index-1 saddle and the lower right index-1 saddle,  respectively, after the bifurcation, i.e. for $1.32 \le c_1 \le 5$ . As we described above, these regions are referred to  as the lower left exit region and the lower right exit region. In addition, we consider that a trajectory has exited through the region of the  upper left index-1 saddle  or the region of the upper right index-1 saddle  if  the trajectories cross the line $y=2.5$ and they have negative $x$-coordinates or positive $x$-coordinates, respectively. If a  trajectory  did not exit from any region in the fixed integration time we consider  that it is trapped in the central area of the caldera. 

\begin{figure}
 \centering
A)\includegraphics[scale=0.55]{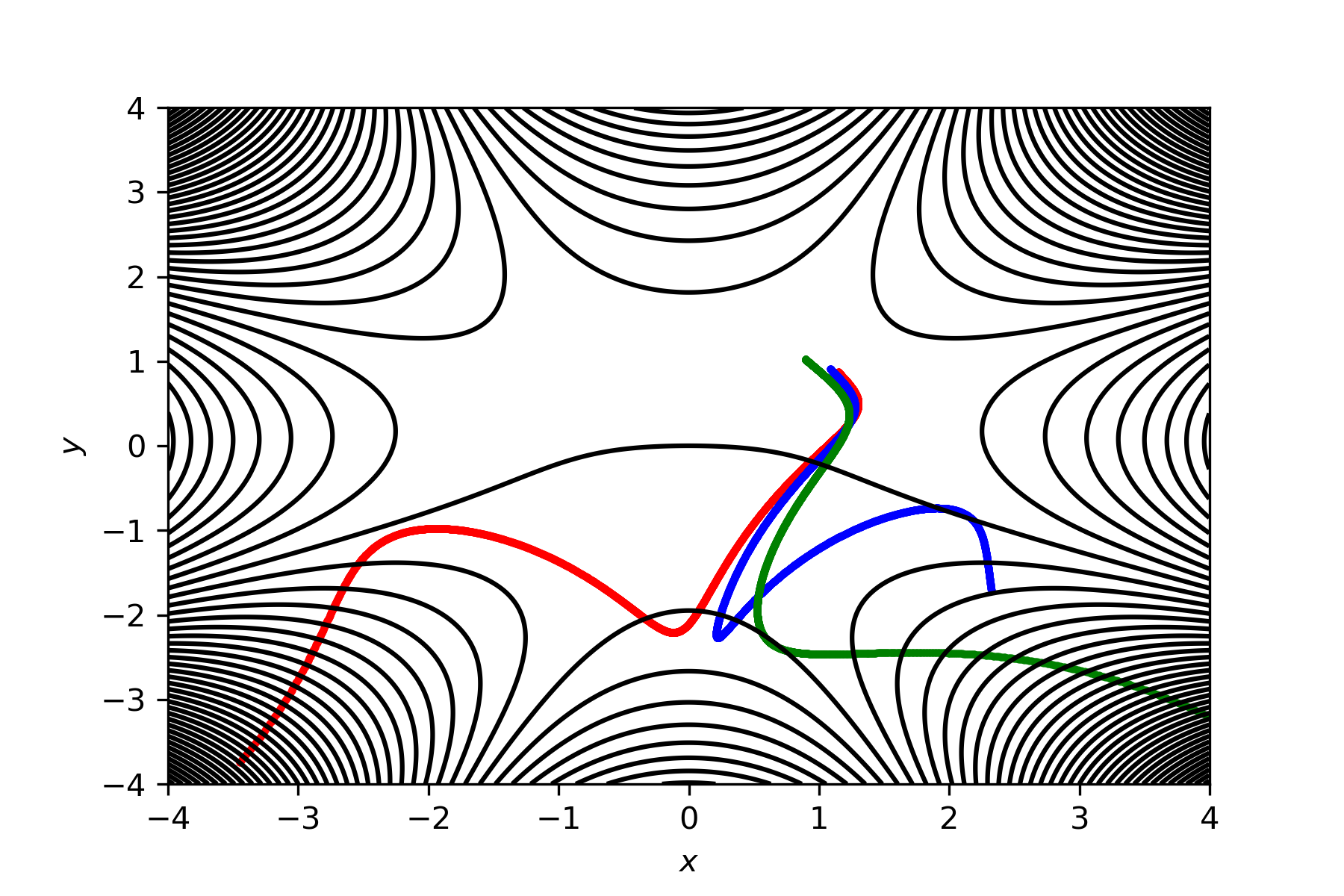}
B)\includegraphics[scale=0.55]{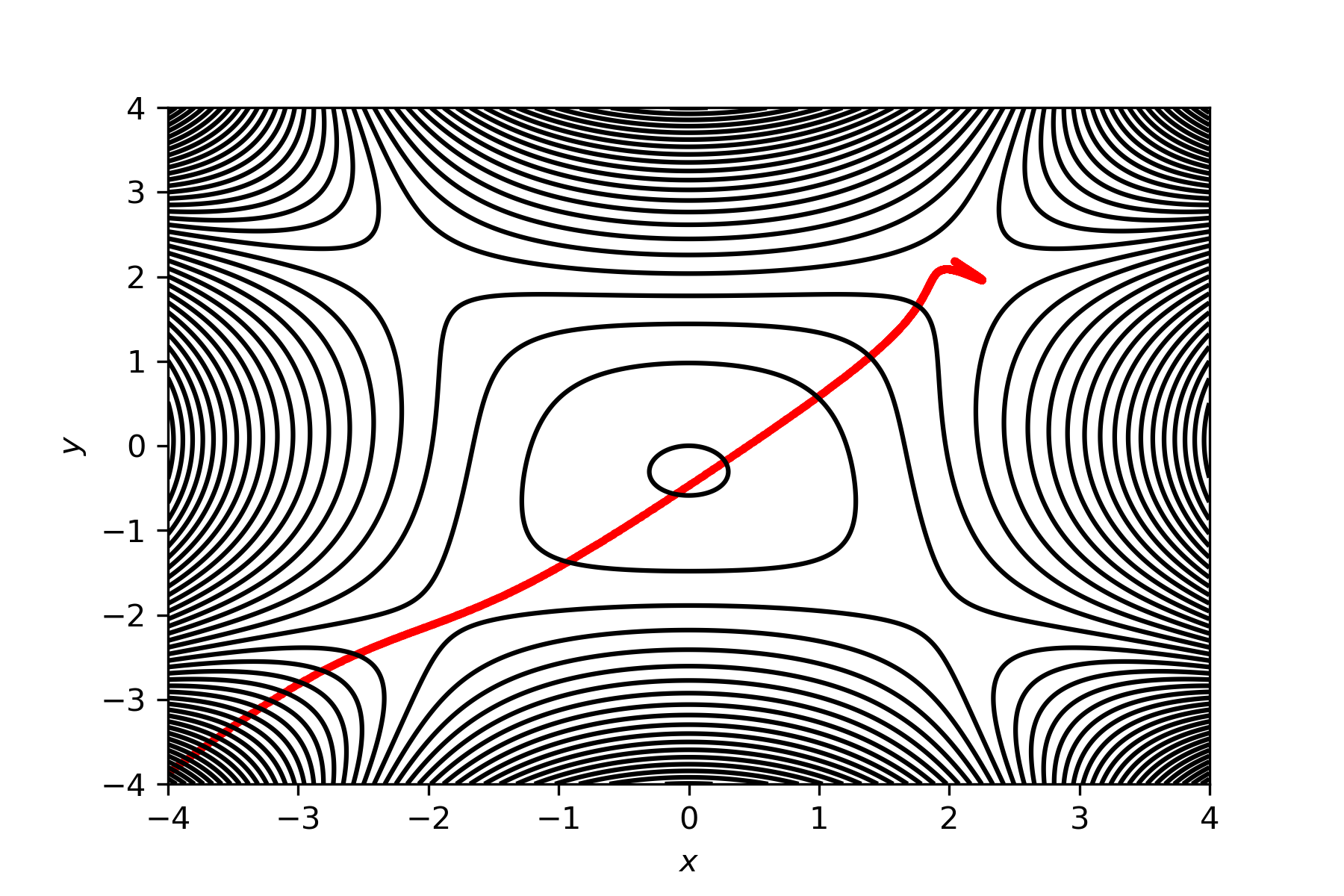}\\
\caption{The contours of the caldera potential and  trajectories in the configuration space that begin from the region of the upper right saddle (for a value of energy 0.5 units above the energy of the higher index-1 saddles): A) for $c_1=0.4,c_2=3$ and $c_3=-0.3$. B)  for $c_1=5,c_2=3$ and $c_3=-0.3$. The trajectories that are trapped or they exit through the region of the lower left saddles  or they exit through the region of the lower right saddle  are depicted by blue, red and green color respectively.}
\label{traj}
\end{figure}

The results of our simulations showed us  that we have three types of behavior of the trajectories that come from the region of the upper right index-1 saddle for $0 \le c_1 \le 1.32$. The first and second  type are represented by the red and green curves, respectively in the panel A of Fig. \ref{traj}. The red and green curves  correspond to  trajectories that exit through the lower left exit region or the lower right exit region, respectively. The third type is represented by the blue line (see the panel A of Fig. \ref{traj}) that corresponds to a trajectory that is trapped in the central region of the caldera.

\begin{figure}
 \centering
\includegraphics[scale=0.5]{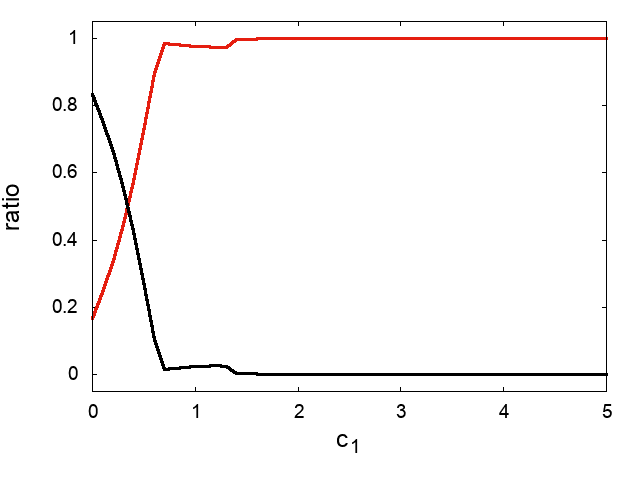}
\caption{The ratio of the trajectories that exit through the lower left exit region (red line) and the ratio of the trajectories that are trapped or they exit through the lower right exit region (black line) versus parameter $c_1$.}
\label{ratio}
\end{figure}

We computed the ratio of the trajectories that exit through the lower left exit region and the ratio of the trajectories that are trapped or they exit through the lower right exit region as the parameter $c_1$ increases. We see in Fig. \ref{ratio} that  initially the ratio of the trajectories that exit through the lower left exit region (red line) is lower than the ratio of the trajectories that are trapped  or exit through the  lower right exit region (black line). Then the ratio of the trajectories that exit through the lower left exit region  increases and it exceeds the ratio of the trajectories that are trapped or they exit through the lower right exit region. This ratio increases until it converges (see the first plateau of the red line in Fig. \ref{ratio})  to a high value (approximately to 0.974). The bifurcation point ($c_1=1.32)$ corresponds to the end of this plateau which is the point where  we have the final increase of the  ratio of the trajectories that exit through the region of the lower left exit region until it converges to one. We observe the exact opposite behavior for the ratio of the trajectories that are trapped or they exit through the region of the lower right exit region (see Fig.\ref{ratio}). 

Furthermore, we see that as $c_1$ increases from the bifurcation value of $1.32$ we have the final convergence of the ratio of the trajectories that exit through the lower left exit region and of the ratio of the trajectories that are trapped or they exit through the lower right exit region to one and zero, respectively. This means that in this case we have only one type of behavior for the trajectories that enter into the caldera from the region of the upper right index-1 saddle. The same happens for the trajectories that enter the caldera from the region of the upper left index-1 saddle as a result of the symmetry of the potential. This type of behavior is that all the trajectories that come from the region of the upper index-1 saddles cross the caldera and they exit though the opposite lower exit region (see for example the panel B of Fig. \ref{traj}). This is the phenomenon of dynamical matching that is found in all previous studies of trajectories in the  symmetric caldera (see \cite{collins2014,katsanikas2018,katsanikas2019,katsanikas2020a,katsanikas2020b}).

\section{Conclusions}
\label{conclusions}

In this paper we investigated the  behavior  of the trajectories that come from the region of the high energy saddles in  caldera potential energy surfaces. We found that the family of the equilibrium points  of the y-axis (the symmetry axis of the caldera) undergoes a pitchfork bifurcation by increasing a parameter of the potential $c_1$.  We observed that for $0 \le c_1 \le 1.32$ the trajectories that enter into the central area of the caldera from the region of the upper index-1 saddle can be trapped in the central area or to follow two different paths to the exit. The one path is to exit through the lower left exit region and the other is where the trajectories  exit through the lower right exit region. The situation for these trajectories is different for $1.32 \le c_1 \le 5$. In this case the trajectories that  enter into the central area of the caldera from the region of the high energy index-1 saddles are not trapped,  and they  have  only one choice  for their exit. This choice is to cross the caldera and to exit through the opposite lower exit region. This behavior is the dynamical matching that has been studied in many papers 
(\cite{collins2014,katsanikas2018,katsanikas2019,katsanikas2020a,katsanikas2020b}). 
Hence, we have shown that a pitchfork bifurcation of critical points of the PES can destroy dynamical matching, even in the case where the PES is symmetric.

\section*{Acknowledgments}

The authors would like to acknowledge the financial support provided by the EPSRC Grant No. EP/P021123/1.
\bibliography{main}

\end{document}